%% file: main.tex
\documentclass[letterpaper, 10 pt, conference]{ieeeconf}  
\IEEEoverridecommandlockouts                             
\input{config/packages}
\input{config/macros}

\title{\LARGE \bf An Innovations-Based Data-Driven Kalman Predictor\\ for Predictive Control}

\author{Mohamed Abdalmoaty and Roy S.\ Smith
\thanks{Authors are with the Automatic Control Laboratory (IfA), Swiss Federal Institute of Technology (ETH Z\"{u}rich), 8092 Z\"{u}rich, Switzerland,
        {\tt\footnotesize \{mabdalmoaty,rsmith\}@control.ee.ethz.ch}}%
}

\begin{document}

\maketitle
\thispagestyle{empty}
\pagestyle{empty}

\begin{abstract}      
A recently developed data-driven Kalman filter requires offline measurement of the process disturbance; a requirement that is often unmet for many practical applications.
We propose a solution that parametrizes the Kalman filter exclusively using measured input and output data. The key idea is to use the innovations form which naturally accounts for the process disturbance and measurement noise into a single orthogonal stochastic process. Unlike process disturbances, the innovations process can be estimated directly from input-output data via a numerically efficient projection step. The performance of the method is demonstrated using a benchmark simulation.
\end{abstract}


\input{sections/introduction}
\input{sections/background}
\input{sections/method}

\input{sections/simulation}
\input{sections/Conclusion}

\bibliographystyle{ieeetr}
\bibliography{library}

\end{document}

%% file: config/packages.tex
\usepackage{graphicx}      
\usepackage{amsmath}
\usepackage{amssymb}
\newtheorem{prop}{Proposition}
\usepackage{comment}
\usepackage{xcolor}
\usepackage{empheq}

%% file: config/macros.tex
\renewcommand{\H}{\mathcal{H}}
\newcommand{\col}{\text{col}}

\newcommand{\nxb}{\bar{n}_x}

\newcommand{\e}{\mathrm{e}}

\newcommand{\xu}{x_u}
\newcommand{\xy}{x_y}
\newcommand{\xuy}{x_{uy}}
\newcommand{\xuyest}{\hat{x}_{uy}}
\newcommand{\xhat}{\hat{x}}
\newcommand{\ub}{\bar{u}} 
 
\newcommand{\upb}{\bar{u}_{\text{p}}} 
\newcommand{\yp}{y_{\text{p}}}

\newcommand{\uf}{u_{\text{f}}}
\newcommand{\ufb}{\bar{u}_{\text{f}}}
\newcommand{\yf}{y_{\text{f}}}

\newcommand{\yfest}{\hat{y}_{\text{f}}}

\newcommand{\yb}{\bar{y}}

\newcommand{\Lupb}{L_{\bar{u}\text{p}}}

\newcommand{\Lyp}{L_{y\text{p}}}

\newcommand{\Lyupb}{L_{y\bar{u}\text{p}}}

\newcommand{\Luf}{L_{u\text{f}}}
\newcommand{\Lufb}{L_{\bar{u}\text{f}}}
\newcommand{\Lyuf}{L_{yu\text{f}}}
\newcommand{\Lyufb}{L_{y\bar{u}\text{f}}}

\newcommand{\Suu}{S_{uu}}
\newcommand{\Suub}{S_{\bar{u}\bar{u}}}
\newcommand{\Suy}{S_{uy}}
\newcommand{\Suyb}{S_{\bar{u}y}}
\newcommand{\Syu}{S_{yu}}
\newcommand{\Syub}{S_{y\bar{u}}}
\newcommand{\Syy}{S_{yy}}

\newcommand{\Auu}{A_{uu}}

\newcommand{\Ayu}{A_{yu}}
\newcommand{\Ayy}{A_{yy}}

\newcommand{\Apred}{A_{\text{p}}}
\newcommand{\Bpred}{B_{\text{p}}}
\newcommand{\Bepred}{B_{\text{ep}}}
\newcommand{\Bupred}{B_{\text{up}}}

\newcommand{\Cpred}{C_{\text{p}}}
\newcommand{\Bpredw}{B_{w\text{p}}}
\newcommand{\Bpredu}{B_{u\text{p}}}

\newcommand{\Euf}{E_{u\text{f}}}
\newcommand{\Exy}{E_{x_y}}
\newcommand{\Exu}{E_{x_u}}

\newcommand{\D}{\mathcal{D}}
\newcommand{\Q}{\mathcal{Q}}
\renewcommand{\P}{\mathcal{P}}

\newcommand{\Buu}{B_{uu}}
\newcommand{\Byu}{B_{yu}}
\newcommand{\Cyu}{C_{yu}}
\newcommand{\Cyy}{C_{yy}}

%% file: sections/introduction.tex
\section{Introduction}
\label{sec:introduction}

Data-driven control methods can be categorized into indirect and direct methods. Indirect methods are based on two steps: first, a model is fitted to identification data, and second, the estimated model is used to design a controller. It is well known that the model with the best data fit, e.g., in the sense of minimizing the one-step ahead prediction error, is not necessarily the best model for the intended control design. Direct methods handle this nuisance by using the data directly for control design. The identification step then becomes implicit as shown in \cite{dorfler2022bridging}.

Data-driven predictive control methods use past input-output trajectories to predict and control the future behavior of systems. Methods based on Willems’ fundamental lemma \cite{willems2005note, de2019formulas} provide linear characterizations of all input-output trajectories consistent with an underlying finite-dimensional linear time-invariant (LTI) system. These have been used to construct predictive controllers directly from data; e.g.,  DeePC \cite{coulson2019data}, $\gamma$-DDPC \cite{breschi2023uncertainty,breschi2023data} and GDPC \cite{lazar2023generalized}. These methods rely on a simulation data-driven model and account for potential disturbances and noises,  in both offline and online data, via regularization. 

Robust predictive control formulations have long addressed disturbances 
explicitly, for instance, through optimization over parameterized feedback policies as in \cite{goulart2006optimization}. Such methods address the future effects of future disturbances.

Kalman filtering methods, in contrast, are based on disturbance models, and account for the effects of past unmeasured disturbances. The recent data-driven formulation of the Kalman filter (KF) in \cite{smith2024data} offers a way to build an optimal filter from measured input-disturbance-output data.  A main limitation of this approach is its requirement that the process disturbance be measurable (offline) for the purpose of constructing the Signal Matrix Model (SMM). Although in some applications this may be possible in a controlled experiment or via a simulation design model, in many other applications it may not be feasible, making direct inclusion of disturbance effects in the data-driven model impossible. 

This paper proposes a general data-driven Kalman filter formulation suitable for realistic cases where the process disturbance is unmeasurable. The key idea is to represent the system in the celebrated \emph{innovations form}, where the process disturbance and any measurement noise are incorporated through the innovations process. The innovations form of linear state-space models comes with many advantages that simplify the analysis and design of Kalman filters as well as subspace identification methods; see e.g.,  \cite{kailath2000linear, lindquist2015linear}.

The innovations process can be estimated (think ``measured") using input–output data by projecting past inputs and outputs onto future outputs. This reconstruction step is a standard preliminary step in several classical subspace identification algorithms, especially when handling closed-loop data \cite{chiuso2007role,van2012subspace}, in addition to more recent data-driven methods 
\cite{breschi2023data,chiuso2025harnessing}. The estimated innovations process can be used directly to build a SMM of the system, including the disturbance dynamics. The resulting framework preserves the advantages of the SMM-based KF while relaxing its requirement for measured disturbance data.



%% file: sections/background.tex
\section{Problem formulation and Background}
\label{sec:background}

\subsection{Problem Formulation}
Consider a discrete-time, linear, time-invariant 
system, with $n_u$ inputs and $n_y$ outputs, 
\vspace{-0.1cm}
\begin{subequations}\label{eq:sys_ss}
\begin{empheq}[left=G\;\empheqlbrace\;]{align}
x(k+1) &= A x(k) + B_u u(k) + B_w w(k), \label{eq:sys_state} \\
\yb(k) &= C x(k) + D u(k) \label{eq:sys_output}\\
y(k) &= \yb(k) + v(k)\nonumber
\end{empheq}
\end{subequations}
where $k$ is a positive integer time index, $x(k) \in \mathbb{R}^{n_x}$ is the state variable,  $u(k) \in \mathbb{R}^{n_u}$, $\yb(k) \in \mathbb{R}^{n_y}$, and $y(k)$ are the control input,  noise-free output and measured output signals, respectively. The signals  $w(k) \in \mathbb{R}^{n_w}$ and $v(k)$ represent process disturbance and measurement noise, respectively. They are modeled as zero-mean white stochastic  processes, mutually uncorrelated,  with finite covariances $\Sigma_w$ and $\Sigma_v$, respectively.  
To simplify the notation, we will use the same symbol to denote a stochastic processes and its realizations.  

In \cite{smith2024data}, Willim's fundamental lemma is used to obtain a data-driven representation of the stochastic system $G$. It was assumed that
a trajectory of realizations $\{u^d(k), w^d(k), y^d(k)\}_{k=1}^N$ is available in the SMM construction step, where the superscript $d$ indicates the offline/historical data used to build the SMM.  In this paper, we remove this assumption and treat $w$ as \emph{unknown and unmeasurable}. Consequently, the SMM cannot be directly constructed in its original form, which motivates a reformulation of the problem. We will assume the availability of an input-output trajectory
\[
\D := \{(u^d(k), y^d(k))\}_{k=1}^{L+N}
\]
where the extra $L$ input-output pairs substitute the knowledge of the unmeasurable disturbance.

Consider a $T$-long online trajectory of inputs and outputs of $G$, divided into $T_p$-long immediate past and $T_f$-long future sub-trajectories with $T= T_f+T_p$. The recursive data-driven prediction problem is then stated as follows.\smallbreak
\begin{quote}
  \emph{Given an input-output trajectory $\D$ and the immediate past inputs and outputs, $u_p(t),  y_p(t)$,  find an estimate of the future outputs $y_f(t)$  in terms of the future \mbox{inputs $u_f(t)$}, conditioned on $u_p(t),  y_p(t)$.}
\end{quote}
\medbreak
For a deterministic $G$, where $w$ and $v$ are identically zero, the above problem becomes a simulation problem with a known data-driven solution \cite{markovsky2008data}. It relies on two conditions. First, the length of the immediate past trajectory $T_p$ must be greater than the lag of the system. This ensures a unique characterization of the underlying state of $G$. In practice, neither the system's lag nor the system's order $n_x$ is known, but an upper bound on the system's order $\bar{n}_x$ is imposed. Second, the input $u^d$ in $\D$ must satisfy a persistency of excitation condition in order to  characterize all trajectories of $G$. When $w$ and $v$ cannot be ignored,  the data-driven simulation is likely to yield inaccurate predictions. To address this, alternative approaches such as subspace predictors \cite{favoreel1999spc} or multi-ARX models \cite{chiuso2025harnessing}, and their variants \cite{verheijen2023handbook} can be used. However, these approaches are static in nature, in the sense that the predictor does not use previous prediction errors. These errors may contain information about the past disturbance that influences the future output and can be exploited by using a predictor with internal dynamics.  

Before presenting our solution, we briefly summarize some relevant existing  results.

\subsection{Data-driven simulation}
For any signal $\zeta$, positive integers $j>k$, let $\zeta_{k:j}$ denote its restriction to the finite  sequence $\{\zeta(k), \dots, \zeta(j)\}$, and define the Hankel matrix 
\[
\begin{aligned}
\mathcal{H}_T&(\zeta_{k:j}) :=\\
&  \begin{bmatrix}
\zeta(k) & \zeta(k+1) & \dots & \zeta(k+M-1)\\
\zeta(k+1) & \zeta(k+2) & \dots & \zeta(k+M+2)\\
\vdots &\vdots &  \ddots & \vdots\\
\zeta(k+T-1) & \zeta(T+1) & \dots & \zeta(j)\\
\end{bmatrix} 
\end{aligned}
\]
with $T\leq j-k+1$ block-rows and the maximal number of columns $M = j-k-T+2$; and the column vectors
\[
    \zeta_p(t) := \begin{bmatrix}
        \zeta(t-T_p-1)\\ \vdots\\ \zeta(t)
    \end{bmatrix}, \quad 
    \zeta_f(t) := \begin{bmatrix}
        \zeta(t+1)\\ \vdots \\ \zeta(t+T_f)
    \end{bmatrix}.
\]
Moreover, define the extended input $\ub(k) = \col\{u(k), w(k)\}$ where $w$ is a known realization of the  process disturbance.

\begin{prop} \label{prop:willims} Consider the system $G$ and assume that $w$ is measurable, and $v(k) = 0$ $\forall k$ so that $y = \yb $. Let $\{(\ub^d(k), y^d(k))\}_{k=1}^N$  be a trajectory of $G$. Then, if the pair $(A,B_u)$ is controllable and the input is persistently exciting of order $T+n_x$, i.e., $\H_{T+n_x}(\ub^d_{1:K})$ has full row rank, then
\begin{itemize}
    \smallbreak
    \item $\{(\ub(k),y(k))\}_{k=1}^T$ is an input-output trajectory of $G$ if and only if there exists $g \in \mathbb{R}^M$ such that
    \begin{equation}\label{eq:trajectory_model}
    \begin{bmatrix}
        \ub\\
        y
    \end{bmatrix} = 
    \begin{bmatrix}
        \H_T(\ub^d_{1:K})\\
        \H_T(y^d_{1:K})
    \end{bmatrix} g,
    \end{equation}
    and $\ub=\col\{\ub(1),\! \dots\!, \ub(T)\}$,  $y=\col\{y(1),\! \dots\!, y(T)\}$.
    \smallbreak
    \item the vector $y_f(t)$ is the unique output trajectory of $G$ with past trajectories $\ub_p(t)$, $y_p(t)$ and input trajectory $\ub_f(t)$, with $T_p \geq n_x$, if and only if there exists a $g \in \mathbb{R}^M$ such that
    \begin{equation}\label{eq:smm}
     \begin{bmatrix}
        \ub_p(t)\\
        \ub_f(t)\\
        y_p(t)\\
        y_f(t)
    \end{bmatrix} = 
    \begin{bmatrix}
        H_{\upb}\\  H_{\ufb}\\  H_{yp}\\ H_{yf} 
    \end{bmatrix} g,
    \end{equation}
    where $H_{\upb}, H_{\ufb}$ are given by the first $T_p$ block-rows and the last $T_f$ block-rows of $\H_T(\bar{u}^d_{1:K})$, respectively, and $H_{yp}, H_{yf}$ are defined similarly using $\H_T(y^d_{1:K})$.
\end{itemize}
\smallbreak
\end{prop}
For details, see \cite{willems2005note} and \cite{de2019formulas} for the first assertion and \cite{markovsky2008data} for the second. Notice that the realizations of $w$ are assumed to be known in all data trajectories. This allows it to be treated the same way as the control input $u$ when applying these results.

The characterization of $G$ in \eqref{eq:smm}, called the SMM, is used directly in DeePC to replace explicit state-space models. Modified versions of it are used by the $\gamma$-DDPC and the GDPC methods, which reduce the dimension of the vector $g$ by removing the null space of the SMM; see \cite{verheijen2023handbook} for an overview of the different methods.

\subsection{Data-based state-space characterization}
Another minimal data-driven characterization of $G$ is given in \cite{smith2024optimal} via a few algebraic operations on the SMM\footnote{Note that in \cite{smith2024optimal}, it is assumed that $D = 0$. Then $y_p$ and $y_f$ are one time step advanced with respect to $\upb, \ufb$, and the Hankel matrices in \eqref{eq:trajectory_model} are formed using $\ub^d_{1:N-1}$ and $y^d_{2:N}$. 
However, in the noise free case,
the rank of the Hankel matrices will sometimes not equal $n_x.$\label{fn:time_shift}} in \eqref{eq:smm}.
The resulting parsimonious SMM has $n_uT + \bar{n}_x$ columns and $g := \col\{x_u, x_y, z\} \in \mathbb{R}^{n_uT + \bar{n}_x}$ is divided into three vectors: $\xu \in \mathbb{R}^{n_uTp}$, $\xy \in \mathbb{R}^{\nxb}$, and $z \in  \mathbb{R}^{n_u T_f}$.  Similarly to the $\gamma$-DDPC schemes \cite{breschi2023data}, the main advantage is the clear separation between the effects of the initial conditions and  $\ufb$ on $\yf$. The variables $\xu$ and $\xy$ are completely determined by $\upb$ and $\yp$, and together in addition to $z$ uniquely determine $\yf$ conditioned on $\uf$. Based on this and an observation that the vector $\xuy(t+1) :=  \col\{\xu(t+1), \xy(t+1)\}$ can be computed in terms of the previous value $\xuy(t)$ and the input $u(t)$, \cite{smith2024data} derived a \textit{data-based} state-space model
\begin{equation}\label{eq:ddstsp_w_v}
\begin{aligned}
\xuy(t+1)   &=  \Apred \, \xuy(t)  +  \Bpredu  u(t) + \Bpredw w(t)\\
y(t)   &=  \Cpred \, \xuy(t) + v(t) . 
\end{aligned}
\end{equation}
where the matrices $\Apred, \Bpredu, \Bpredw$, and $\Cpred$ are completely determined using $\{u^d(k), w^d(k), y^d(k)\}_{k=1}^K$ through {algebraic operations} on the SMM.  This is in essence similar to classical subspace identification algorithms when measured disturbances are included in the data matrices. Yet, the Willem's lemma route taken there allows for particular state-space coordinates. A standard stationary KF \cite{kailath2000linear} can be designed to estimate the state $\xuy$, representing the initial conditions, effectively handling the process disturbance and measurement noise in the online data trajectory. Let $\xuyest(t)$ denote the \textit{filtered} state at time $t$. Then, a predictor of $y_f$ in terms of $u_f$ is obtained using 
\[
\yfest(t)  = 
E_{x_{uy}} \xuyest(t) +  \Euf \uf(t)
\]
where, again, the matrices in the last equation are obtained via algebraic operations. In other words, all future output are predicted simultaneously via the SMM and the KF is only used for filtering and estimating $\xuy(t)$.

%% file: sections/method.tex
\section{Data-Driven Kalman Prediction}
\label{sec:method}

\subsection{Innovations Form Representation}

An alternative representation of $G$ is given by the 
stationary  innovations form,
\begin{equation}\label{eq:innov_state}
\begin{aligned}
\xhat(k+1) &= A \xhat(k) + B_u u(k) + K e(k), \\
y(k) &= C \hat{x}(k) + D u(k) + e(k),
\end{aligned}
\end{equation}
This is the predictor form of the KF associated with \eqref{eq:sys_ss} where the unique innovations process, $e(k) \in \mathbb{R}^{n_y}$, is a zero-mean white stochastic process with covariance matrix $\Lambda$.  It represents the optimal one-step-ahead linear prediction error of $y(k)$, and encapsulates the stochastic effects of both $w(\ell)$ and $v(\ell)$ $\forall \ell\leq k$. The equivalence between the two stationary models \eqref{eq:sys_ss} and \eqref{eq:innov_state} is in the sense that $y$ has exactly the same second-order statistics. Notice that the state space matrices $(A,B,C, D)$ are the same in both models. 
For a finite time horizon, $ k \in \{1, \dots, \ell\}$, the relationship between the realizations of the innovations and $w, v$ depends on the predicted initial state $\hat{x}(1)$ of \eqref{eq:innov_state}. Suppose that $\hat{x}(1) = 0$ (mean-square optimal when the prior distribution is centered); then by propagating the state prediction errors, $x(k)-\hat{x}(k)$, and using the second row in \eqref{eq:innov_state}, we see that
\[
\begin{aligned}
e(\ell) &= C(A-KC)^{\ell-1} x(1) \\
&+ C\sum_{k=1}^{\ell-1}(A-KC)^{N-k-1}( B_w w(k) - K v(k)).
\end{aligned}
\]
The innovations form has several known advantages, with the main one being its causal invertibility. That is, the roles of $e$ and $y$ as input/output of the system in \eqref{eq:innov_state} can be interchanged. 
This causal invertibility is in sharp contrast to the original form in \eqref{eq:sys_ss}, where it is clearly impossible to compute $x(1)$, and $\{ (w(k),v(k))\}_{k =1}^\ell$ from knowledge of $\{ (u(k),y(k))\}_{k=1}^\ell$. 
The innovations form has been used to derive closed-loop subspace identification algorithms and subspace predictors; see, e.g., \cite{favoreel1999spc,qin2003closed,chiuso2007role} and \cite{breschi2023data, chiuso2025harnessing}.

\subsection{Innovations-based SMM}
Because \eqref{eq:innov_state} provides a complete characterization of $G$, where the output is unchanged but  the input signal is now modified to $\ub := \col\{u(k), e(k)\} \in \mathbb{R}^{n_u+n_y}$, the following natural result follows immediately.
\smallbreak
\begin{prop}
    The assertions of Proposition \ref{prop:willims} hold without the assumption that $v(k)=0\; \forall k$, if in all the data trajectories used there, the process disturbance $\{w(k)\}$ is replaced by the innovation process $\{e(k)\}$.
\end{prop}
\smallbreak

This means that we can use the innovations-based SMM
\begin{equation}\label{eq:innov-SMM}
    \begin{bmatrix}
        \H_T(\bar{u}^d_{1:K})\\
        \H_T(y^d_{1:K})
    \end{bmatrix} 
\end{equation}
to characterize all innovations-input-output trajectories of \eqref{eq:innov_state}. Notice that the measured output (including the measurement noise, $v$) is used in this innovations-based SMM, unlike in \eqref{eq:trajectory_model}. This is because $e$ accounts for the influence of both $w$ and $v$.
We can now obtain a parsimonious version of \eqref{eq:innov-SMM}.
\smallbreak
\begin{prop}
Consider the characterization in \eqref{eq:smm} with $\ub = \col\{u(k), e(k)\}$ and where $y$ is the measured output. An equivalent characterization is given as
\begin{equation}\label{eq:parsimonous_smm}
\begin{bmatrix}  
\upb \\ \yp \\ \ufb \\ \yf
\end{bmatrix}
 = 
	\begin{bmatrix} \Lupb   & 0        & 0  \\
	                \Lyupb & \Lyp & 0 \\
	                \Suub & \Suyb & \Lufb \\
	                \Syub & \Syy & \Lyufb 
    \end{bmatrix}
	  \begin{bmatrix}  
                \xu \\ \xy \\ z 
    \end{bmatrix},
\end{equation}	
where $\xu \in \mathbb{R}^{n_uTp}$, $\xy \in \mathbb{R}^{\nxb}$, and $z \in  \mathbb{R}^{n_u T_f}$.  
\end{prop}
\smallbreak
The matrices $\Lupb$, $\Lyupb$, and $\Lyp$ are obtained using the $LQ$ decomposition of $\col\{H_{\bar{u}p}, H_{yp}\}$. The matrices $\Lufb$ and $\Lyufb$ are obtained using the $LQ$ decomposition of the projection of $\col\{H_{\bar{u}f}, H_{yf}\}$ into the null space of $\col\{H_{\bar{u}p}, H_{yp}\}$. The four remaining matrices account for the influence of $\upb$ and $\yp$ on $\bar{u}_f$ and $y_f$ via $x_u$ and $x_y$. The reader is referred to \cite{smith2024optimal}  for a complete derivation.

\subsection{An innovations-based data-driven KF}
We now derive a recursive predictor based on \eqref{eq:parsimonous_smm}. Unlike \cite{smith2024data}, we consider a direct feed-through term in the model and therefore the control input and output signals are not time shifted with respect to each other (see footnote \ref{fn:time_shift}).  Observe that, even when $D=0$,  \eqref{eq:innov_state} has a direct feed-through  from $e$, which is treated as an input when constructing the SMM.

Define the block upper shift matrix $S_n = S \otimes I_n$, where $[S]_{i,j} = \delta_{i, j-1}$, and the block selection matrix \mbox{$J_n = \e_{T_p}\otimes I_n$} in which $\e_{T_p} \in \mathbb{R}^{T_p}$  with all entries equal zero except the last entry equals one. 
Then
\[
\begin{aligned}
    \ub_p(t) &=  S_{n_{\ub}}\ub_p(t-1) +  J_{n_{\ub}} \ub(t)\\
    y_p(t) &=  S_{n_y}y_p(t-1) +  J_{n_y} y(t)\\
\end{aligned}
\]
From this and the first two block-rows in $\eqref{eq:parsimonous_smm}$,
we get
\[
\begin{aligned}
    \xu(t) &=  \Auu \xu(t-1) +  \Buu \ub(t),
\end{aligned}
\]
where
\[
\begin{aligned}
\Auu &:=\Lupb S_{n_{\ub}}\Lupb^{-1},\\
\Buu &:= \Lupb^{-1} J_{n_{\ub}},
\end{aligned}
\]
and
\[
\begin{aligned}
    \Lyp\xy(t) &= -  \Lyupb\xu(t) + S_{n_y}\Lyupb\xu(t-1)  \\
     & \quad +  S_{n_y}\Lyp\xy(t-1) +  J_{n_y} y(t)\\
     & = (S_{n_y}\Lyupb -  \Lyupb\Auu)\xu(t-1) \\
     & \quad  - \Lyupb\Buu \ub(t)+  S_{n_y}\Lyp\xy(t-1) +  J_{n_y} y(t).
\end{aligned}
\]
By multiplying both sides by $\Pi := \begin{bmatrix} I_{T_p - 1}&0\end{bmatrix} \otimes I_{n_y}$,
\[
\begin{aligned}
    \xy(t)  & = \Ayu\xu(t-1) + \Ayy \xy(t-1) + \Byu\ub(t) 
\end{aligned}
\]
where
\[
\begin{aligned}
\Ayu &:= \hphantom{-} \Phi \Pi (S_{n_y}\Lyupb -  \Lyupb\Auu) ,\\
\Ayy &:=\hphantom{-} \Phi \Pi\; S_{n_y}\Lyp,\\
\Byu &:= -\Phi \Pi \; \Lyupb\Buu,
\end{aligned}
\]
and $\Phi = ( \Lyp^\top \Pi^\top\Pi \Lyp)^{-1}\Lyp^\top \Pi^\top$ is the pseudo-inverse of $ \Pi \Lyp$. Therefore, we arrive at the following relations
\[
\begin{aligned}
    \xuy(t) &=  \Apred \xuy(t-1) +  \Bpred \ub(t),\\
    y(t)  & = \Cpred \xuy(t)
\end{aligned}
\]
where
\[
\begin{aligned}
    \Apred = \begin{bmatrix} 
    \Auu & 0 \\
    \Ayu & \Ayy
    \end{bmatrix}, \quad
    \Bpred = \begin{bmatrix} 
    \Buu\\
    \Byu
    \end{bmatrix}, \quad
    \Cpred = \begin{bmatrix} 
    \Cyu &  \Cyy
    \end{bmatrix}
\end{aligned}
\]
and $\Cyu = J_{n_y}^\top \Lyupb, \;\Cyy = J_{n_y}^\top \Lyp$. \smallbreak

To get a standard state-space model, re-index the states by defining the vector $\xuy^+(t+1) = \xuy(t)$ and write $\Bpred =\begin{bmatrix} \Bupred & \Bepred \end{bmatrix}$ where $\Bepred$ has $n_y$ columns. The final model is
\begin{equation}\label{eq:ddstsp_e}
\begin{aligned}
    \xuy^+(t+1) &=  \Apred \xuy^+(t) +  \Bupred u(t) + \Bepred e(t),\\
    y(t)  & = \Cpred\Apred \xuy^+(t) + \Cpred\Bupred u(t) + \Cpred\Bepred e(t)
\end{aligned}
\end{equation}
Note the direct feed-through terms in the output equation arising from $\eqref{eq:innov_state}$.
\smallbreak

A standard stationary KF can be designed to estimate the state $\xuy^+$. This provides an efficient way of handling disturbances and measurement noise in the online data trajectory. For completeness, the recursions of the KF are included here. The model in \eqref{eq:ddstsp_e} has a particular disturbance/noise correlation structure; The Kalman gain is given by \cite{kailath2000linear}
\[
K_{\text{pred}} = (\Apred P \Cpred^\top + \Lambda_{12})(\Cpred P \Cpred^\top + \Lambda_{2})^{-1}
\]
where $P$ is the positive definite solution of the Riccati equation
\[
\begin{aligned}
P = \Apred P \Apred^\top &+ \Lambda_1 - (\Apred P \Cpred^\top + \Lambda_{12})\\
&\times(\Cpred P \Cpred^\top + \Lambda_{2})^{-1} (\Apred P \Cpred^\top + \Lambda_{12})^\top
\end{aligned}
\]
and  $\Lambda_1 =  \Bepred\Lambda \Bepred^\top, \;  \Lambda_2 =  \Cpred\Lambda_1\Cpred^\top, \; \Lambda_{12} = \Lambda_1 \Cpred$. The KF update equation is
\[
\begin{aligned}
\hat{x}^+_{uy}(t+1|t) &=\Apred \hat{x}^+_{uy}(t|t-1) + \Bupred u(t)\\
& + K_{\text{pred}} (y(t) - \Cpred (\Apred\hat{x}^+_{uy}(t|t-1) + \Bupred u(t)))
\end{aligned}
\]
At time $t$, the predicted state, $\hat{x}^+_{uy}(t+1|t)$, coincides with  the estimate of  $\hat{x}_{uy}(t)$ at time $t$. Then, a predictor of $y_f$ in terms of $u_f$ is obtained using the last two block-rows of the SMM \eqref{eq:parsimonous_smm} as
\[
\yfest(t)  = 
\begin{bmatrix}\Exu & \Exy \end{bmatrix} \xuyest(t) +  \Euf \uf(t)
\]
where
\[
\begin{aligned}
    \Exu &:= \Syu - \Lyuf\Luf^{-1}\Suu \\
    \Exy &:=  \Suy - \Lyuf\Luf^{-1}\Suy\\
    \Euf &:= [\Lyuf\Luf^{-1}]_{:n_uT_f}.
\end{aligned}
\]
In other words, the SMM generates all future output predictions simultaneously and the KF is only used for filtering  and estimating $\xuy(t)$.

\subsection{``Measuring" the innovations $e^d$}
Since $\D$ contains values of $u^d$ and $y^d$ only, we need to find an estimate of $e^d$ before building the innovations-based SMM. This can be done by using the first $L$ samples in $\D$ and approximating the impulse-response of the inverse of $\eqref{eq:innov_state}$ where $y$ is the input and $e$ is the output. A similar innovation estimation step is used in several classical subspace identification algorithms as well as recent data-driven methods (\cite{chiuso2007role,van2013closed,breschi2023data,chiuso2025harnessing}). \smallbreak

From \eqref{eq:innov_state}, we can write
\[
\xhat(k+1) = (A-KC) \xhat(k) + B_u u(k) + K y(k),
\]
and by propagating the state forward in time, we obtain the following set of matrix equations that relate the inputs, outputs, and innovations via the system matrices $(A,B,C,K)$
\begin{equation}\label{eq:ARX}
    \begin{aligned}
        \H_{1}(y^d_{L+1:L+N})&  =  \P\; \H_L(\zeta^d_{1:L+N-1})   + \Q \; \hat{x}(1) \\
        & \quad\; + \H_1(e^d_{L+1:L+N}),
    \end{aligned}
\end{equation}
where $\zeta^d(k) := \col\{ u^d(k),  y^d(k)\}$,
\begin{align*}
\bar{A} &= A - KC, \qquad    \bar{B} = \begin{bmatrix} B_u & K \end{bmatrix},\\
\P &= \begin{bmatrix} C\bar{A}^{L-1}\bar{B} &  C\bar{A}^{L-2}\bar{B} & \dots  &\bar{B} \end{bmatrix}, \text{ and }\\
\Q &= \begin{bmatrix}
 C\bar{A}^L  \dots & C\bar{A}^{L+N-1}
\end{bmatrix}.
\end{align*}
Assuming $L$ is large enough, the effect of $\hat{x}(1)$ in \eqref{eq:ARX} is negligible, and an estimate of the innovations sequence is obtained via the $LQ$ decomposition
\begin{equation*}
    \begin{bmatrix}
        \H_L(\zeta^d_{1:L+N-1})\\
        \H_1(y^d_{L+1:L+N})
    \end{bmatrix}      =
    \begin{bmatrix}
        L_{11}  & 0  \\
        L_{21}  & L_{22}
    \end{bmatrix}  
    \begin{bmatrix}
        Q^\top_1\\
        Q^\top_2
    \end{bmatrix},
\end{equation*}
where $L_{11}$ and $L_{22}$ are square matrices with dimensions $(n_u+n_y)L$ and $n_y$, respectively, and $Q^\top_2 \in \mathbb{R}^{n_y \times N}$, as
\[
\begin{aligned}
    {\H_1(\hat{e}^d_{L+1:L+N})} &= \begin{bmatrix} \hat{e}_{L+1}^d & \hat{e}_{L+2}^d & \dots & \hat{e}_{L+N}^d\end{bmatrix}\\
    & := L_{22}Q_2^\top  \in \mathbb{R}^{n_y \times N}
\end{aligned}
\]
Notice that only the innovations starting at $L+1$ to $L+N$ are estimated. The innovations-based SMM can then be built using the estimated innovations, $\hat{e}^d_{L+1:L+N}$, the inputs and outputs, $u^d_{L+1:L+N}$, $y^d_{L+1:L+N}$,  as
\begin{equation}\label{eq:innov-SMM-est}
    \begin{bmatrix}
        \H_T(\bar{u}^d_{L+1:L+N})\\
        \H_T(y^d_{L+1:L+N})
    \end{bmatrix} 
\end{equation}
in which  $\bar{u}^d(k) = \col\{u^d(k), \hat{e}^d(k) \}$, and the first $L$ samples are not used (because $e_{1:L}^d$ are not estimated). The covariance matrix of the innovations, $\Lambda$, can be estimated using the sample covariance of $\hat{e}^d_{L+1:L+N}$.
\smallbreak

The estimate  $\hat{e}^d_{L+1:L+N}$ can be seen as a noise-corrupted measurement of the true innovation via a soft sensor; namely $\hat{e}^d_{L+1:L+N}= e^d_{L+1:L+N}  + \varepsilon_{L+1:L+N}$.   The nature of the errors $\varepsilon$ depends on $N$ and the number of samples $L$ (past horizon), which is a design parameter. The decay rate of the entries of $\mathcal{P}$ and $\mathcal{Q}$ depends on the spectral radius of $\bar{A}$, which is partly dictated by the Kalman gain $K$. When the covariance $\Sigma_v$ is small compared to $\Sigma_w$, the Kalman gain is large, leading to faster decay even when the modes of $A$ are slow. 

Observe that the above innovations sequence estimate corresponds to the residuals of a vector auto-regressive model of order $L$, which is fitted implicitly. For small $L, N$, the bias-variance trade-off in estimating the innovations is more prominent and $L$ should be chosen carefully. e.g., using an Akaike information criterion or a cross-validation method. This initial projection step provides a noise averaging effect. Indeed, the resulting innovation-based SMM \eqref{eq:innov-SMM-est} is ``consistent", i.e., the errors $\varepsilon$ vanish asymptotically as $N$ grows unbounded if $L$ also grows with $N$ with an appropriate rate; see \cite{kuersteiner2005automatic} and the discussion in  \cite{chiuso2007role}. 

\begin{figure*}[!t]
\centering
    \includegraphics[width=0.85\linewidth]{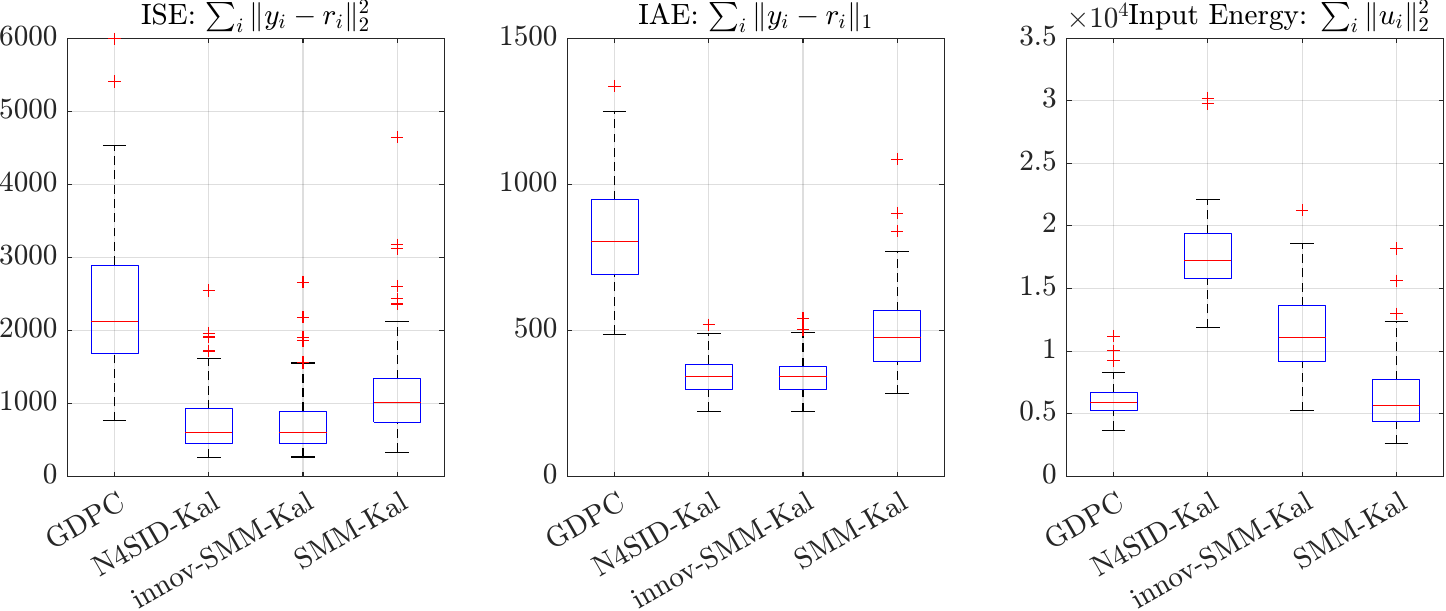}\vspace{-0.25cm}
    \caption{Performance indices boxplots. The performance of innov-SMM-KF is a bit better than SMM-Kal that uses measured $w$ to construct the SMM.}
    \label{fig:boxplots}\vspace{-0.1cm}
\end{figure*}

%% file: sections/simulation.tex
\section{Simulation Example}
\label{sec:simulation}
To illustrate the performance of the proposed method, we repeat the simulation example from \cite{smith2024data} where a linearized continuous-time model of the longitudinal dynamics of the Boeing 747 aircraft is considered. This is also the same model used in \cite{verheijen2023handbook}, but disturbances were not considered there. It has the following  state-space matrices
\begin{gather*}
A = \begin{bmatrix}
    -0.003 & 0.039 & 0 & -0.322\\
    -0.065 & -0.319 & 7.74 & 0\\
    0.02 & -0.101 & -0.429 & 0\\
    0 & 0 & 1 & 0
    \end{bmatrix},\\
B_u = \begin{bmatrix}0.010 & 1\\
-0.18 & -0.04\\
-1.16 & 0.598\\
0 & 0\end{bmatrix}, \qquad C= \begin{bmatrix}
    1 & 0 & 0& 0\\
    0 & -1 & 0 & 7.74
\end{bmatrix},
\end{gather*}
where the two control inputs are the throttle, $u_1$, and the elevator angle, $u_2$ [deg], and the two outputs are the longitudinal velocity, $y_1$ [ft/s], and the climb rate $y_2$ [ft/s]. The four states are the longitudinal velocity, $x_1$ [ft/s], the downward velocity, $x_2$ [ft/s], the pitch angular velocity, $x_4$ [deg].
Disturbances are modeled using the Dryden gust model with 
\[
B_w = \begin{bmatrix}
    -1 & 0\\
    0 & -1\\
    0 & 0\\
    0 & 0
\end{bmatrix} \begin{bmatrix} 
w_{h,\text{gust}}\\
w_{v,\text{gust}}
\end{bmatrix},
\]
where the gust spectra are given by the transfer functions
\[
\begin{aligned}
    w_{h,\text{gust}} &= \sigma_{u,\text{gust}} \sqrt{\frac{2L_u}{V\pi}} \frac{1}{1+ (L_u/V)s}w_1,\\
    w_{v,\text{gust}} &= \sigma_{v,\text{gust}} \sqrt{\frac{2L_v}{V\pi}} \frac{1+(2\sqrt{3}L_v/V)s}{(1+ 2(L_v/V)s)^2}w_2
\end{aligned}
\]
in which $\sigma_{u,\text{gust}} = \sigma_{v,\text{gust}} = 10$, and turbulence length scales $L_u = 1750$ ft., $L_v = L_u/2$, and $s$ is the Laplace variable. The model is discretized with zero-order hold and a uniform  sample period of 0.1 s. The discrete-time measurement and process noise are Gaussian random variables with covariances $\Sigma_v = 0.25^2 I_2$ and $\Sigma_w = I_2$.
\smallbreak

For the prediction and control problems, the following values are used. The number of data points in $\mathcal{D}$ is $N = 2500$ and $L = 150$ ($L$ was not tuned). The past and future horizons are $T_p = 30$ and $T_f = 20$. The performance of the method is evaluated via Monte Carlo simulations, where 100 different independent realizations of the process disturbance and measurement noise are used.  In addition, a reference step change is simulated at $t = 3$ s. For each simulation, the \textit{noisy data} is used to build a SMM \eqref{eq:innov-SMM}, and the closed-loop predictive control simulations have both (online) measurement noise and unmeasured process disturbance. The following four methods are compared
\begin{itemize}
    \item { \bf \textit{GDPC}:} A regularized data-driven predictive control \cite{lazar2023generalized}. This method is taken as a representative of the class of regularization-based data-driven prediction methods. 
    \item {\bf  \textit{N4SID-Kal}:} Identified $7^\text{th}$ order subspace model (using N4SID) + a Kalman filter with MPC. The identification data contains only inputs and outputs (i.e., based on $\mathcal{D}$). 
    \item {\bf  \textit{SMM-Kal}:} The SMM-based Kalman filter from \cite{smith2024data}, using measurements of $w$ to build the SMM.
    \item {\bf  \textit{innov-SMM-Kal}:} The approach proposed in this paper: an innovation-based SMM-Kalman filter.
\end{itemize}

\begin{figure}
\centering
    \includegraphics[width=\linewidth]{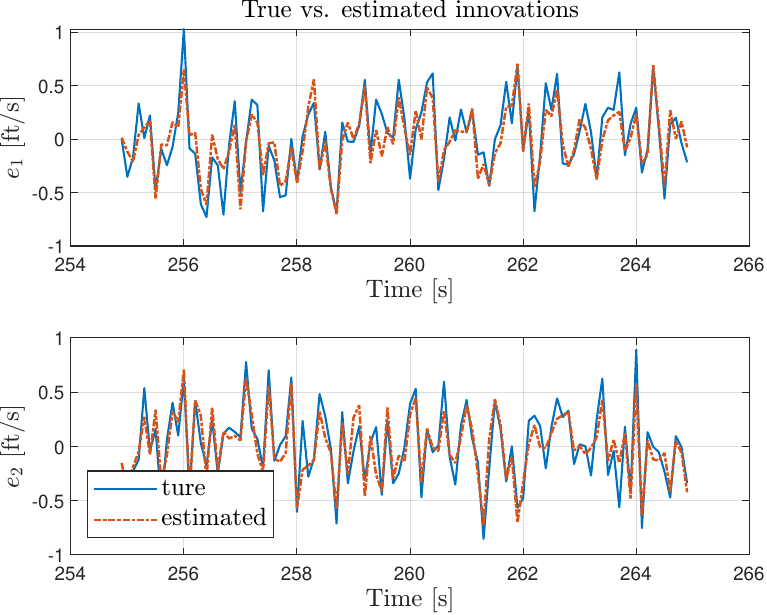}\vspace{-0.25cm}
    \caption{True versus estimated innovations sequence with $L = 150$ samples. For clarity, only the tail of the 2500 samples trajectory is shown.}\vspace{-0.5cm}
    \label{fig:estimated_innov}
\end{figure}

Figure \ref{fig:boxplots} shows the boxplots of several performance indices (these are the same indices used in \cite{verheijen2023handbook, smith2024data}). The performance of the innovations-based SMM KF is a bit better than the performance of the SMM-based KF, which uses measured disturbance trajectory to build the SMM. This is because the SMM of the latter has noise in both $w^d$ and $y^d$, while the SMM of the former has an averaging effect with noise only in $\hat{e}^d$. This demonstrates the effectiveness of the proposed approach. Notice that innovations-based SMM KF relies only on $\D$ without assuming that the disturbance $w$ is accessible for measurement. The true and estimated innovations  are shown in Figure \ref{fig:estimated_innov}. As demonstrated in \cite{smith2024data}, other  data-driven methods, represented here by GDPC, do not perform as well; this is expected because they do not have any characterization of the dynamics of the disturbance. The performance of the N4SID-Kal method is comparable to that of innov-SMM-Kal, while using more input energy. It estimates a state-space model in innovations form (i.e., with disturbance model) using the true order (7 in this example).

%% file: sections/conclusion.tex
\section{Conclusion}
\label{sec:conclusion}

This work addresses a practical limitation of a recent data-driven Kalman filter (KF): the requirement for an offline measurement of the process disturbance. For many real-world applications where this disturbance is unknown or unmeasured, this requirement makes the method impractical. To overcome this, we introduced a solution that parametrizes the data-driven KF exclusively using measured input and output data. The central insight is to leverage the innovation form of the system, which naturally encodes the effects of both the process disturbance and the measurement noise into a single orthogonal stochastic process—the innovation process. Crucially, this innovation process can be estimated directly from the data via a numerically efficient projection step, bypassing the need to measure the process disturbance separately. As demonstrated through a benchmark simulation, this method maintains strong performance and significantly broadens the applicability of the data-driven KF to a wider range of practical control problems.

%% file: library.bib
@article{willems2005note,
  title={A note on persistency of excitation},
  author={Willems, Jan C and Rapisarda, Paolo and Markovsky, Ivan and De Moor, Bart LM},
  journal={Systems \& Control Letters},
  volume={54},
  number={4},
  pages={325--329},
  year={2005},
  publisher={Elsevier}
}

@article{de2019formulas,
  title={Formulas for data-driven control: Stabilization, optimality, and robustness},
  author={De Persis, Claudio and Tesi, Pietro},
  journal={IEEE Transactions on Automatic Control},
  volume={65},
  number={3},
  pages={909--924},
  year={2019},
  publisher={IEEE}
}

@inproceedings{coulson2019data,
  title={Data-enabled predictive control: In the shallows of the {DeePC}},
  author={Coulson, Jeremy and Lygeros, John and D{\"o}rfler, Florian},
  booktitle={2019 18th European control conference (ECC)},
  pages={307--312},
  year={2019},
  organization={IEEE}
}

@article{breschi2023uncertainty,
  title={Uncertainty-aware data-driven predictive control in a stochastic setting},
  author={Breschi, Valentina and Fabris, Marco and Formentin, Simone and Chiuso, Alessandro},
  journal={IFAC-PapersOnLine},
  volume={56},
  number={2},
  pages={10083--10088},
  year={2023},
  publisher={Elsevier}
}

@article{breschi2023data,
  title={Data-driven predictive control in a stochastic setting: a unified framework},
  author={Breschi, Valentina and Chiuso, Alessandro and Formentin, Simone},
  journal={Automatica},
  volume={152},
  pages={110961},
  year={2023},
  publisher={Elsevier}
}

@article{lazar2023generalized,
  title={Generalized data--driven predictive control: Merging subspace and {H}ankel predictors},
  author={Lazar, M and Verheijen, PCN},
  journal={Mathematics},
  volume={11},
  number={9},
  pages={2216},
  year={2023},
  publisher={MDPI}
}

@inproceedings{smith2024data,
  title={Data-driven formulation of the {K}alman filter and its application to predictive control},
  author={Smith, Roy S and Abdalmoaty, Mohamed and Yin, Mingzhou},
  booktitle={2024 IEEE 63rd Conference on Decision and Control (CDC)},
  pages={2633--2639},
  year={2024},
  organization={IEEE}
}

@article{dorfler2022bridging,
  title={Bridging direct and indirect data-driven control formulations via regularizations and relaxations},
  author={D{\"o}rfler, Florian and Coulson, Jeremy and Markovsky, Ivan},
  journal={IEEE Transactions on Automatic Control},
  volume={68},
  number={2},
  pages={883--897},
  year={2022},
  publisher={IEEE}
}

@article{markovsky2008data,
  title={Data-driven simulation and control},
  author={Markovsky, Ivan and Rapisarda, Paolo},
  journal={International Journal of Control},
  volume={81},
  number={12},
  pages={1946--1959},
  year={2008},
  publisher={Taylor \& Francis}
}

@article{verheijen2023handbook,
  title={Handbook of linear data-driven predictive control: Theory, implementation and design},
  author={Verheijen, PCN and Breschi, Valentina and Lazar, Mircea},
  journal={Annual Reviews in Control},
  volume={56},
  pages={100914},
  year={2023},
  publisher={Elsevier}
}

@inproceedings{smith2024optimal,
  title={Optimal data-driven prediction and predictive control using signal matrix models},
  author={Smith, Roy S and Abdalmoaty, Mohamed and Yin, Mingzhou},
  booktitle={2024 IEEE 63rd Conference on Decision and Control (CDC)},
  pages={6767--6773},
  year={2024},
  organization={IEEE}
}

@book{kailath2000linear,
  title={Linear Estimation},
  author={Kailath, T. and Sayed, A.H. and Hassibi, B.},
  isbn={9780130224644},
  lccn={99047033},
  year={2000},
  publisher={Prentice Hall}
}

@article{lindquist2015linear,
  title={Linear stochastic systems},
  author={Lindquist, Anders and Picci, Giogio},
  journal={Series in Contemporary Mathematics},
  volume={1},
  pages={26},
  year={2015},
  publisher={Springer}
}

@book{van2012subspace,
  title={Subspace identification for linear systems: Theory—Implementation—Applications},
  author={Van Overschee, Peter and De Moor, BL0888},
  year={2012},
  publisher={Springer Science \& Business Media}
}

@article{favoreel1999spc,
  title={{SPC}: Subspace predictive control},
  author={Favoreel, Wouter and De Moor, Bart and Gevers, Michel},
  journal={IFAC Proceedings Volumes},
  volume={32},
  number={2},
  pages={4004--4009},
  year={1999},
  publisher={Elsevier}
}

@article{qin2003closed,
  title={Closed-loop subspace identification with innovation estimation},
  author={Qin, S Joe and Ljung, Lennart},
  journal={IFAC Proceedings Volumes},
  volume={36},
  number={16},
  pages={861--866},
  year={2003},
  publisher={Elsevier}
}

@article{chiuso2007role,
  title={The role of vector autoregressive modeling in predictor-based subspace identification},
  author={Chiuso, Alessandro},
  journal={Automatica},
  volume={43},
  number={6},
  pages={1034--1048},
  year={2007},
  publisher={Elsevier}
}

@article{kuersteiner2005automatic,
  title={Automatic inference for infinite order vector autoregressions},
  author={Kuersteiner, Guido M},
  journal={Econometric Theory},
  volume={21},
  number={1},
  pages={85--115},
  year={2005},
  publisher={Cambridge University Press}
}

@article{verhaegen1992dewilde,
author = {Michel Verhagen and Patrik Dewilde},
title = {Subspace model identification Part 1. The output-error state-space model identification class of algorithms},
journal = {International Journal of Control},
volume = {56},
number = {5},
pages = {1187--1210},
year = {1992},
publisher = {Taylor \& Francis},
}

@article{van2013closed,
  title={Closed-loop subspace identification methods: an overview},
  author={Van der Veen, Gijs and van Wingerden, Jan-Willem and Bergamasco, Marco and Lovera, Marco and Verhaegen, Michel},
  journal={IET Control Theory \& Applications},
  volume={7},
  number={10},
  pages={1339--1358},
  year={2013},
  publisher={Wiley Online Library}
}

@article{goulart2006optimization,
  title={Optimization over state feedback policies for robust control with constraints},
  author={Goulart, Paul J and Kerrigan, Eric C and Maciejowski, Jan M},
  journal={Automatica},
  volume={42},
  number={4},
  pages={523--533},
  year={2006},
  publisher={Elsevier}
}

@article{chiuso2025harnessing,
  title={Harnessing uncertainty for a separation principle in direct data-driven predictive control},
  author={Chiuso, Alessandro and Fabris, Marco and Breschi, Valentina and Formentin, Simone},
  journal={Automatica},
  volume={173},
  pages={112070},
  year={2025},
  publisher={Elsevier}
}
